\documentclass[a4paper,UKenglish,cleveref, autoref]{lipics-v2019}

\hideLIPIcs

\title{Simple Analysis of Sparse, Sign-Consistent JL} 

\author{Meena Jagadeesan}{Harvard University, Cambridge, MA, USA }{mjagadeesan@college.harvard.edu}{}{Supported in part by a Harvard PRISE fellowship, Herchel-Smith Fellowship, and an REU supplement to NSF IIS-1447471.}
\authorrunning{M. Jagadeesan}

\Copyright{Meena Jagadeesan}



\supplement{}

\acknowledgements{I would like to thank Prof. Jelani Nelson for advising this project.}

\DeclareMathOperator{\Bin}{Bin}
\def\eps{\varepsilon}
\newcommand{\norm}[1]{\left\| #1 \right\|}               

\usepackage{url}

\nolinenumbers 

\EventEditors{Dimitris Achlioptas and L\'{a}szl\'{o} A. V\'{e}gh}
\EventNoEds{2}
\EventLongTitle{Approximation, Randomization, and Combinatorial Optimization. Algorithms and Techniques (APPROX/RANDOM 2019)}
\EventShortTitle{\mbox{\scriptsize APPROX/RANDOM 2019}}
\EventAcronym{APPROX/RANDOM}
\EventYear{2019}
\EventDate{September 20--22, 2019}
\EventLocation{Massachusetts Institute of Technology, Cambridge, MA, USA}
\EventLogo{}
\SeriesVolume{145}
\ArticleNo{61}

\begin{document}

\maketitle

\begin{abstract}
Allen-Zhu, Gelashvili, Micali, and Shavit construct a sparse, sign-consistent Johnson-Lindenstrauss distribution, and prove that this distribution yields an essentially optimal dimension for the correct choice of sparsity. However, their analysis of the upper bound on the dimension and sparsity requires a complicated combinatorial graph-based argument similar to Kane and Nelson's analysis of sparse JL. We present a simple, combinatorics-free analysis of sparse, sign-consistent JL that yields the same dimension and sparsity upper bounds as the original analysis. Our analysis also yields dimension/sparsity tradeoffs, which were not previously known. 

As with previous proofs in this area, our analysis is based on applying Markov's inequality to the $p$th moment of an error term that can be expressed as a quadratic form of Rademacher variables. Interestingly, we show that, unlike in previous work in the area, the traditionally used Hanson-Wright bound is \textit{not} strong enough to yield our desired result. Indeed, although the Hanson-Wright bound is known to be optimal for gaussian degree-2 chaos, it was already shown to be suboptimal for Rademachers. Surprisingly, we are able to show a simple moment bound for quadratic forms of Rademachers that is sufficiently tight to achieve our desired result, which given the ubiquity of moment and tail bounds in theoretical computer science, is likely to be of broader interest.
\end{abstract}

\section{Introduction}
In many modern algorithms that process high dimensional data, it is beneficial to preprocess the data through a dimensionality reduction scheme that preserves the geometry of the data. Dimensionality reduction schemes have been applied in streaming algorithms \cite{Streaming} as well as algorithms for numerical linear algebra \cite{NumLinAlg}, feature hashing \cite{Weinberger}, graph sparsification \cite{GSparse}, and many other areas. The geometry-preserving objective can be expressed mathematically as follows. The goal is to construct a probability distribution $\mathcal{A}$ over $m \times n$ real matrices that satisfies the following condition for any $x \in \mathbb{R}^n$:
\begin{equation}
\label{JLcondition}
\mathbb{P}_{A \in \mathcal{A}} [(1 - \epsilon)||x||_2 \le ||Ax||_2 \le (1 + \epsilon)||x||_2] > 1-\delta.
\end{equation} An upper bound on the dimension $m$ achievable by a probability distribution $\mathcal{A}$ that satisfies $\eqref{JLcondition}$ is given in the following lemma, which is a central result in the area of dimensionality reduction: 
\begin{lemma}[Johnson-Lindenstrauss \cite{JL}]
\label{RegularJL}
For any positive integer $n$ and parameters $0 < \epsilon, \delta < 1$, there exists a probability distribution $\mathcal{A}$ over $m \times n$ real matrices with $m = \Theta(\epsilon^{-2} \log(1/\delta))$ that satisfies \eqref{JLcondition}.  
\end{lemma}
\noindent The optimality of the dimension $m$ achieved by Lemma~$\ref{RegularJL}$ was recently proven in \cite{OptimalityJL1, OptimalityJL2}. 

For many applications of dimensionality reduction schemes, it can be useful to consider probability distributions over sparse matrices in order to speed up the projection time. Here, sparsity refers to the constraint that there are a small number of nonzero entries in each column. In this context, Kane and Nelson \cite{KN12} constructed a sparse JL distribution, improving the work of Achlioptas \cite{Achlioptas} and Dasgupta et al. \cite{DKS10}, and proved the following: 
\begin{theorem}[Sparse JL \cite{KN12}]
\label{SparseJL}
For any positive integer $n$ and $0 < \epsilon, \delta < 1$, there exists a probability distribution $\mathcal{A}$ over $m \times n$ real matrices with $m = \Theta(\epsilon^{-2} \log(1/\delta))$ and sparsity $s = \Theta(\epsilon^{-1} \log(1/\delta))$ that satisfies \eqref{JLcondition}.  
\end{theorem}
\noindent Notice that this probability distribution, even with its sparsity guarantee, achieves the same dimension as Lemma~$\ref{RegularJL}$. The proof of Theorem~$\ref{SparseJL}$ presented in \cite{KN12} involved complicated combinatorics; however, Cohen, Jayram, and Nelson \cite{NelsonNotes} recently constructed two simple, combinatorics-free proofs of this result. The first approach, which is most relevant to the approach taken in this paper, used the Hanson-Wright bound on moments of quadratic forms. An analysis similar to the second approach can be recovered by specializing the analysis of Cohen \cite{Cohen} for sparse oblivious subspace embeddings to the case of ``1-dimensional
subspaces.'' In fact, though this recovered analysis is more complex, it has the advantage of yielding dimension-sparsity tradeoffs that were not produced through any of the previous approaches: for $B \ge e$, the sparsity $s$ can be set to $\Theta(\epsilon^{-1} \log_B(1/\delta))$ if $m$ is set to $\Theta(B\epsilon^{-2}\log(1/\delta))$, enabling a $\log B$ factor reduction in sparsity at the expense of a $B$ factor gain in dimension. 

\subsection*{JL with sign-consistency constraints}
Neuroscience-based constraints give rise to the additional condition of sign-consistency on the matrices in the probability distribution. Sign-consistency refers to the constraint that the nonzero entries of each column are either all positive or all negative. The relevance of dimensionality reduction schemes in neuroscience is described in a survey by Ganguli and Sompolinsky \cite{GS12}. In convergent pathways in the brain, information stored in a massive number of neurons is compressed into a small number of neurons, and nonetheless the ability to perform the relevant computations is preserved. Modeling this information compression scheme requires a hypothesis regarding what properties of the original information must be accurately transmitted to the receiving neurons. A plausible minimum requirement is that convergent pathways preserve the similarity structure of neuronal representations at the source area.\footnote{This requirement is based on the experimental evidence that semantically similar objects in higher perceptual or association areas in the brain elicit similar neural activity patterns \cite{Sim1} and on the hypothesis that the similarity structure of the neural code is the basis of our ability to categorize objects and generalize responses to new objects \cite{Sim2}.} 

It remains to select the appropriate mathematical measure of similarity. The candidate similarity measure considered in \cite{GS12} is vector inner product, which conveniently gives rise to a model based on the JL distribution.\footnote{It is not difficult to see that for vectors $x$ and $y$ in the $\ell_2$ unit ball, a $(1+\eps)$-approximation of $\norm{x}_2$, $\norm{y}_2$, and $\norm{x-y}_2$ implies an additive error $\Theta(\epsilon)$ approximation of the inner product $\langle x, y \rangle$.} Suppose there are $n$ ``input'' neurons at a source area and $m$ ``output'' neurons at a target area. In this framework, the information at the input neurons is represented as a vector in $\mathbb{R}^n$, the synaptic connections to output neurons are represented as a $m \times n$ matrix (with $(i,j)$th entry corresponding to the strength of the connection between input neuron $j$ and output neuron $i$), and the information received by the output neurons is represented as a vector in $\mathbb{R}^m$. The similarity measure between two vectors $v, w$ of neural information being taken to be $\langle v, w \rangle$ motivates modeling a synaptic connectivity matrix as a random $m \times n$ matrix drawn from a probability distribution that satisfies \eqref{JLcondition}. Certain constraints on synaptic connectivity matrices arise from the biological limitations of neurons: the matrices must be \textit{sparse} since a neuron is only connected to a small number (e.g. a few thousand) of postsynaptic neurons and \textit{sign-consistent} since a neuron is usually purely excitatory or purely inhibitory. 

This biological setting motivates the mathematical question: what is the optimal dimension and sparsity that can be achieved by a probability distribution over sparse, sign-consistent matrices that satisfies \eqref{JLcondition}? Allen-Zhu, Gelashvili, Micali, and Shavit \cite{Original} constructed a sparse, sign-consistent JL distribution\footnote{Related mathematical work includes, in addition to sparse JL \cite{KN12}, a construction of a dense, sign-consistent JL distribution \cite{Rajan, Dense2}.} and proved the following:
\begin{theorem}[Sparse, sign-consistent JL \cite{Original}]
\label{OriginalMainResult}
For every $\eps > 0$, and $0 < \delta < 1/e$, there exists a probability distribution $\mathcal{A}$ over $m \times n$ real, sign-consistent matrices with $m = \Theta(\eps^{-2}\log^2(1/\delta))$ and sparsity $s = \Theta(\eps^{-1}\log(1/\delta))$ that satisfies \eqref{JLcondition}. 
\end{theorem}
\noindent In \cite{Original}, it was also proven that the additional  $\log(1/\delta)$ factor on $m$ is essentially necessary: namely, any distribution over sign-consistent matrices satisfying $\eqref{JLcondition}$ requires $m = \tilde{\Omega}(\eps^{-2}\log(1/\delta)\min(\log (1/\delta), \log n))$. Thus, the dimension in Theorem~$\ref{OriginalMainResult}$ is essentially optimal. However, in order to achieve this upper bound on $m$, the proof presented in \cite{Original} involved complicated combinatorics even more delicate than in the analysis of sparse JL in \cite{KN12}. 

We present a simpler, combinatorics-free proof of  Theorem~$\ref{OriginalMainResult}$. Our analysis also yields dimension/sparsity tradeoffs, which were not previously known.\footnote{In Appendix A, we point out the limiting lemma in the combinatorial analysis in \cite{Original}, which prevents dimension-sparsity tradeoffs from being attainable through this approach, due to an assumption that is implicitly used in the analysis. For sparse JL, it is similarly not known how to obtain these tradeoffs via the combinatorial approach of \cite{KN12}.} We prove the following:  
\begin{theorem}
\label{OurMainResult}
For every $\eps > 0$, $0 < \delta < 1$, and $e \le B \le \frac{1}{\delta}$, there exists a probability distribution $\mathcal{A}$ over $m \times n$ real, sign-consistent matrices with $m = \Theta(B\eps^{-2}\log^2_B(1/\delta))$ and sparsity $s = \Theta(\eps^{-1}\log_B(1/\delta))$ that satisfies \eqref{JLcondition}. 
\end{theorem}
\noindent Notice Theorem~$\ref{OriginalMainResult}$ is recovered if $B = e$. For larger $B$ values, Theorem~$\ref{OurMainResult}$ enables a $\log B$ factor reduction in sparsity at the cost of a $B /\log^2 B$ factor gain in dimension. 

To contextualize our tradeoff in Theorem \ref{OurMainResult}, recall that the upper bounds on sparse (non-sign-consistent) JL dimension-sparsity tradeoffs by Cohen \cite{Cohen} take a similar form, allowing a $\log B$ factor reduction in $s$ for a $B$ factor gain in $m$. Moreover, in a recent follow-up work \cite{FeatureHashing}, we show lower bounds that indicate that the standard choice of sparse JL construction requires an exponential factor gain in dimension for a given reduction in sparsity, demonstrating that Cohen's dimension-sparsity tradeoffs are essentially tight.\footnote{More specifically, it follows from \cite{Cohen} and \cite{FeatureHashing} that $m$ is exactly $\min(\text{poly}(B) \epsilon^{-2} \log(1/\delta), 2 \epsilon^{-2} /\delta)$ for the standard choice of sparse JL construction (uniformly choosing $s$ nonzero entries per column).} Due to the structural similarity between sparse JL and sparse, sign-consistent JL, we believe this provides indication that our tradeoffs in Theorem \ref{OurMainResult} could be tight for this construction in many regimes.\footnote{An interesting direction for future work could be to build upon the ideas in the follow-up work \cite{FeatureHashing} to show lower bounds on the dimension-sparsity tradeoffs for this sparse, sign-consistent JL construction.}

\subsection*{Proof Techniques}
As in \cite{Original, KN12, NelsonNotes}, our analysis is based on applying Markov's inequality to the $p$th moment of an error term. Like in the first combinatorics-free analysis of sparse JL in \cite{NelsonNotes}, we express this error term as a quadratic form of Rademachers (uniform $\pm 1$ random variables), and our analysis then boils down to analyzing the moments of this quadratic form. While the analysis in \cite{NelsonNotes} achieves the optimal dimension for sparse JL using an upper bound on the moments of quadratic forms of subgaussians due to Hanson and Wright \cite{HansonWright}, we give a counterexample in Section 3.2 that shows that the Hanson-Wright bound is too loose in the sign-consistent setting to result in the optimal dimension. Since the Hanson-Wright bound is tight for quadratic forms of gaussians, we thus require a separate treatment of quadratic forms of Rademachers. 

We construct a simple bound on moments of quadratic forms of Rademachers that, unlike the Hanson-Wright bound, is sufficiently tight in our setting to prove Theorem~$\ref{OurMainResult}$. Our bound borrows some of the ideas from Lata{\l}a's tight bound on the moments of quadratic forms of Rademachers \cite{LatalaChaos}. Although our bound is much weaker than the bound in \cite{LatalaChaos} in the general case, it has the advantage of providing a greater degree of simplicity by consisting of easier-to-analyze terms; this simplicity is critical since our quadratic form coefficients are themselves random variables. The crux is that while the bound in \cite{LatalaChaos} is focused on obtaining tight estimates for quadratic forms with scalar coefficients, our bound is much more tractable for quadratic forms with random variable coefficients. As a result, our bound enables a simple proof of Theorem \ref{OurMainResult}, while retaining the necessary precision to recover the optimal dimension. 

We build upon these ideas in our recent follow-up work \cite{FeatureHashing}, where the Hanson-Wright bound also turns out to be too loose. The work studies sparse JL performance in feature hashing and considers the restricted set of vectors with small $\ell_\infty$-to-$\ell_2$ norm ratio, continuing a line of work \cite{Weinberger, DKS10, Derandomized, DKT17, KN12, FKL}. The main result is a tight tradeoff between $\ell_\infty$-to-$\ell_2$ norm ratio and $\epsilon$, $\delta$, $s$, and $m$, and the lower bounds on dimension-sparsity tradeoffs mentioned before are shown as a corollary.
Similar to this work, the proof boils down to a tight bound on $p$th moment of an error term, and it also turns out that the Hanson-Wright bound is too loose here. The work solves this issue by building upon ideas from this work, utilizing a separate treatment of Rademachers that is tractable for random variable coefficients. While the analysis in \cite{FeatureHashing} does not use the exact quadratic form bound presented here, it uses intuition and generalizations of the moment bounding techniques presented in this work. 

\subsection{Notation}
The main building blocks for our expressions are the following two types of random variables: Rademacher variables, which are uniform $\pm 1$ random variables, and Bernoulli random variables, which have support $\left\{0,1 \right\}$. For any random variable $X$ and value $p \ge 1$, we use the notation $\norm{X}_p$ to denote the $p$-norm $\left(\mathbb{E}[|X|^p]\right)^{1/p}$, where $\mathbb{E}$ denotes the expectation. Similarly, for any random variable $X$ and value $p \ge 1$ and any event $E$, we use the notation $\norm{X \mid E}_p$ to denote the conditional $p$-norm $\left(\mathbb{E}[|X|^p \mid E ]\right)^{1/p}$, which is equivalent to the $p$-norm of the random variable $(X \mid E)$. We use the following notation to discuss certain asymptotics: given two scalar quantities $Q_1$ and $Q_2$ that are functions of some parameters, we use the notation $Q_1 \simeq Q_2$ to denote that there exist positive universal constants $C_1 \le C_2$ such that $C_1Q_2 \le Q_1 \le C_2Q_2$, and we use the notation $Q_1 \lesssim Q_2$ to denote that there exists a positive universal constant $C$ such that $Q_1 \le CQ_2$. 

\subsection{A digression on Rademachers versus gaussians}
The concept that drives our moment bound can be illustrated in the linear form setting. Suppose $\sigma_1, \sigma_2 \ldots, \sigma_n$ are i.i.d Rademachers, $x= [x_1, \ldots, x_n]$ is a vector in $\mathbb{R}^n$ such that $|x_1| \ge |x_2| \ge \ldots \ge |x_n|$, and $2 \le p \le n$. The Khintchine inequality, which is tight for linear forms of gaussians, yields the $\ell_2$-norm bound $\norm{\sum_{i=1}^n \sigma_ix_i}_{p} \lesssim \sqrt{p}\norm{x}_2$. However, this bound \textit{cannot} be a tight bound on $\norm{\sum_{i=1}^n \sigma_ix_i}_{p}$ for the following reason: As $p \rightarrow \infty$, the quantity $\sqrt{p}\norm{x}_2$ goes to infinity, while for any $p \ge 1$, the quantity $\norm{\sum_{i=1}^n \sigma_ix_i}_{p}$ is bounded by $\norm{x}_1$. Surprisingly, a result due to Hitczenko \cite{OriginalNorm} indicates that the tight bound is actually the following combination of the $\ell_2$ and $\ell_1$ norm bounds:  
\[\norm{\sum_{i=1}^n \sigma_ix_i}_{p} \simeq \sum_{i=1}^p |x_i| + \sqrt{p}\sqrt{\sum_{i>p} x_i^2}.\]
In this bound, the ``big'' terms (i.e. terms involving $x_1$, $x_2$, \ldots, $x_p$) are handled with an $\ell_1$-norm bound, while the remaining terms are approximated as gaussians and bounded with an $\ell_2$-norm bound. 

A similar complication arises when the Hanson-Wright bound on quadratic forms of subgaussians is applied to Rademachers. Let $\sigma$ be a $d$-dimensional vector of independent Rademachers, and let $A = (a_{k,l})$ be a symmetric $d \times d$ matrix with zero diagonal. The Hanson-Wright bound \cite{HansonWright}, which is tight for gaussians, states for any $p \ge 1$,
\[\norm{\sigma^TA\sigma}_{p} \lesssim \sqrt{p} \sqrt{\sum_{k=1}^d \sum_{l=1}^d a_{k,l}^2} + p \left(\sup_{\norm{y}_2 = 1} |y^TAy| \right).\] Similar to the linear form setting, this bound \textit{can't} be a tight bound on $\norm{\sigma^T A \sigma}_{p}$ for the following reason: As $p \rightarrow \infty$, the quantity $\sqrt{p} \sqrt{\sum_{k=1}^d \sum_{l=1}^d a_{k,l}^2}$ goes to $\infty$, while for any $p \ge 1$, the quantity $\norm{\sigma^TA\sigma}_{p}$ is bounded by the entrywise $\ell_1$-norm $\sum_{k=1}^d \sum_{l=1}^d |a_{k,l}|$.

Our quadratic form bound is based on a degree-2 analog of Hitczenko's observation. We analogously handle the ``big'' terms with an $\ell_1$-norm bound and bound the remaining terms by approximating some of the Rademachers by gaussians. From this, we obtain a combination of $\ell_2$ and $\ell_1$ norm bounds, similar to the linear form setting. Our simple bound has the surprising feature that it yields tighter guarantees than the Hanson-Wright bound yields for our error term. While our bound is weaker than Lata{\l}a's tight bound \cite{LatalaChaos}  on the moments of quadratic forms of Rademachers in the general case, it provides a greater degree of simplicity: our bound avoids an operator-norm-like term in Lata{\l}a's bound that is especially difficult to analyze when $A$ is a random matrix, as is the case in this setting. Moreover, our bound still retains the necessary precision to recover the optimal dimension for sparse, sign-consistent JL. 

Although our final analysis follows a style that this is perhaps less well-known within the TCS community, in the end, it is quite simple, relying only on our quadratic form bound coupled with a few standard tricks such as repeated use of triangle inequalities on $|| \cdot||_p$ norms and standard moment bounds involving the binomial distribution. For this reason, we believe that it is likely to be of interest in other theoretical computer science settings involving moments or tail bounds of Rademacher forms. 

\subsection{Outline for the rest of the paper}
In Section 2, we describe the construction and analysis of \cite{Original} for sparse, sign-consistent JL. In Section 3, we present the combinatorics-free approach in \cite{NelsonNotes} for sparse JL that uses the Hanson-Wright bound, and we discuss why this approach does not yield the optimal dimension in the sign-consistent setting. In Section 4, we derive our bound on the moments of quadratic forms of Rademachers and use this bound to construct a combinatorics-free proof of Theorem~$\ref{OurMainResult}$. 

\section{Existing Analysis for Sparse, Sign-Consistent JL}
In Section 2.1, we describe how to construct the probability distribution of sparse, sign-consistent matrices analyzed in Theorem~$\ref{OriginalMainResult}$. In Section 2.2, we briefly describe the combinatorial proof of Theorem~$\ref{OriginalMainResult}$ presented in \cite{Original}. 
\subsection{Construction of Sparse, Sign-Consistent JL}
The entries of a matrix $A \in \mathcal{A}$ are generated as follows.\footnote{See Appendix \ref{appendix:probspace} for a formal construction of the probability space.}
Let $A_{i,j} = \eta_{i,j} \sigma_j / \sqrt{s}$ where $\left\{\sigma_i \right\}_{i \in [n]}$ and $\left\{ \eta_{r,i} \right\}_{r \in [m], i \in [n]}$ are defined as follows:
\begin{itemize}
\item The families $\left\{\sigma_i \right\}_{i \in [n]}$ and $\left\{ \eta_{r,i} \right\}_{r \in [m], i \in [n]}$ are independent from each other.
\item The variables $\left\{\sigma_i \right\}_{i \in [n]}$ are i.i.d Rademachers.
\item The variables $\left\{ \eta_{r,i} \right\}_{r \in [m], i \in [n]}$ are identically distributed Bernoulli random variables with expectation $s/m$.
\item The $\left\{ \eta_{r,i} \right\}_{r \in [m], i \in [n]}$ are independent across columns but not independent within each column. For every column $1 \le i \le n$, it holds that $\sum_{r=1}^m \eta_{r,i} = s$. For every subset $S \subseteq [m]$ and every column $1 \le i \le n$, it holds that $\mathbb{E} \left[\prod_{r \in S} \eta_{r, i}\right] \le \prod_{r \in S} \mathbb{E} [\eta_{r, i}]$. (One common definition of  $\left\{\eta_{r,i}\right\}_{r \in [m], i \in [n]}$ that satisfies these conditions is the distribution defined by uniformly choosing exactly $s$ of these variables per column to be a $1$.)
\end{itemize}
For every $x \in \mathbb{R}^n$ such that $\norm{x}_2 = 1$, we need to analyze an error term, which for this construction is the following random variable:
\[Z := \norm{Ax}^2_2 - 1 = \frac{1}{s} \sum_{i \neq j} \sum_{r=1}^m \eta_{r,i} \eta_{r,j} \sigma_i \sigma_j x_i x_j.\] Proving that $\mathcal{A}$ satisfies $\eqref{JLcondition}$ boils down to proving that $\mathbb{P}_{\eta, \sigma} [|Z| > \epsilon] < \delta$. The main technique to prove this tail bound is the moment method. Bounding a large moment of $Z$ is useful since it follows from Markov's inequality that
\[\mathbb{P}_{\eta, \sigma} [|Z| > \epsilon] = \mathbb{P}_{\eta, \sigma} [|Z|^p > \epsilon^p] < \frac{\mathbb{E} [|Z|^p]}{\epsilon^p}.\] The usual approach, used in the analyses in \cite{Original, KN12, NelsonNotes} as well as in our analysis, is to take $p = \Theta(\log(1/\delta))$ to be an even integer and analyze the $p$-norm $\norm{Z}_{p}$ of the error term. 

\subsection{Discussion of the combinatorial analysis of \cite{Original}}
In the analysis in \cite{Original}, a complicated combinatorial argument was used to prove the following lemma, from which Theorem~$\ref{OriginalMainResult}$ follows:
\begin{lemma}[\cite{Original}]
\label{mainbound}
If $s^2 \le m$ and $p < s$, then $\norm{Z}_{p} \lesssim \frac{p}{s}$.
\end{lemma}
\noindent The argument in \cite{Original} to prove Lemma~$\ref{mainbound}$ was based on expanding $\mathbb{E}[Z^p]$ into a polynomial with $\approx n^{2p}$ terms, establishing a correspondence between the monomials and the multigraphs, and then doing combinatorics to analyze the resulting sum. The approach of mapping monomials to graphs is commonly used in analyzing the eigenvalue spectrum of random matrices \cite{Wigner, Kom} and was also used in \cite{KN12} to analyze sparse JL. The analysis in \cite{Original} borrowed some methods from the analysis in \cite{KN12}; however, the additional correlations between the Rademachers imposed by sign-consistency forced the analysis in \cite{Original} to require more delicate manipulations at several stages of the computation.

The expression to be analyzed was $s^p \mathbb{E}[Z^p]$, which was written as:
\[\sum_{i_1, \ldots, i_p, j_1, \ldots, j_p \in [n], \\ i_1 \neq j_1, \ldots, i_p \neq j_p} \left(\prod_{u=1}^p x_{i_u}x_{j_u}\right) \left(\mathbb{E}_{\sigma} \prod_{u=1}^p \sigma_{i_u}\sigma_{j_u} \right) \left(\mathbb{E}_{\eta} \prod_{u=1}^t \sum_{r=1}^m \eta_{r, i_u} \eta_{r, j_u} \right).\]
\noindent After layers of computation, it was shown that 
\[s^p \mathbb{E}[Z^p] \le e^p \sum_{v=2}^p \sum_{G \in \mathcal{G}_{v,p}} \left((1/p^p) \prod_{q=1}^v \sqrt{d_q}^{d_q} \right) \sum_{r_1, \ldots, r_p \in [m]} \prod_{i=1}^w (s/m)^{v_i}\] where $\mathcal{G}_{v,p}$ is a set of directed multigraphs with $v$ labeled vertices and $t$ labeled edges, where $d_q$ is the total degree of vertex $q \in [v]$ in a graph $\mathcal{G}_{v,p}$, and where $w$ and $v_1, \ldots, v_w$ are defined by $G$ and the edge colorings $r_1, \ldots, r_t$. The problem then boiled down to carefully enumerating the graphs in $\mathcal{G}_{v,p}$ in six stages and analyzing the resulting expression.
\section{Discussion of Combinatorics-Free Approaches}
The main ingredient of the first combinatorics-free approach for sparse JL presented in \cite{NelsonNotes} is the Hanson-Wright bound on the moments of quadratic forms of subgaussians. In Section 3.1, we discuss the approach in \cite{NelsonNotes}. In Section 3.2, we discuss why this approach, if applied to sparse, sign-consistent JL, fails to yield the optimal dimension. 
\subsection{Hanson-Wright approach for sparse JL in \cite{NelsonNotes}}
The relevant random variable for sparse JL is 
\[Z' = ||Ax||^2 - 1 = \frac{1}{s} \sum_{r=1}^m  \sum_{i \neq j} \eta_{r,i} \eta_{r,j} \sigma_{r,i} \sigma_{r,j} x_i x_j\] where the $n$ independent Rademachers $\left\{\sigma_i \right\}_{i \in [n]}$ from the sign-consistent case are replaced by the $mn$ independent Rademachers $\left\{\sigma_{r,i} \right\}_{i \in [n], r \in [m]}$. The main idea in \cite{NelsonNotes} was to view $Z'$ as a quadratic form $\frac{1}{s}\sigma^TA\sigma$. Here, $\sigma$ is a $mn$-dimensional vector of independent Rademachers and $A = (A_{k,l})$ is a symmetric, zero diagonal, block diagonal $mn \times mn$ matrix with $m$ blocks of size $n \times n$, where the $(i,j)$th entry (for $i \neq j$) of the $r$th block is $\eta_{r,i} \eta_{r,j} x_i x_j$. The quantity $\norm{\sigma^T A \sigma}_{p}$ was analyzed using the Hanson-Wright bound.
In order to bound $\norm{\sigma^T A \sigma}_p$, since $A$ is a random matrix whose entries depend on the $\eta$ values, an expectation had to be taken over $\eta$ in the expression given by the Hanson-Wright bound. This resulted in the following:
\begin{equation}
\label{sparseHW}
\norm{\sigma^TA\sigma}_{p} \lesssim \norm{\sqrt{p} \sqrt{\sum_{k=1}^{mn} \sum_{l=1}^{mn} A_{k,l}^2} + p \sup_{\norm{y}_2 = 1} |y^TAy|}_{p}.
\end{equation} \noindent The remainder of the analysis boiled down to bounding the RHS of \eqref{sparseHW}, and it successfully recovered Theorem \ref{SparseJL}.

\subsection{Failure of the Hanson-Wright approach for sparse, sign-consistent JL}
The Hanson-Wright-based approach for sparse JL in \cite{NelsonNotes} cannot be applied to the sign-consistent case to obtain a tight bound on $\norm{Z}_{p}$. The loss arises from the fact that while the Hanson-Wright bound is tight for quadratic forms of gaussians, it is not guaranteed to be tight for quadratic forms of Rademachers. As discussed in Section 1.2, when $p \rightarrow \infty$, the Hanson-Wright bound goes to $\infty$, while $||\sigma^TA\sigma||_p$ can be bounded by the entrywise $\ell_1$ norm of the matrix $A$. Although approximating the error term Rademachers by gaussians happened to be sufficiently tight for sparse JL, this loss results in a suboptimal dimension for sparse, sign-consistent JL.\footnote{The difference results from the correlations between the signs resulting in more ``tightly packed'' coefficients in the error term quadratic form in the sign-consistent case.} We give a counterexample, i.e. a vector $x$, that shows that the Hanson-Wright bound is too loose to give the optimal dimension (when $\left\{\eta_{r,i}\right\}_{r \in [m], i \in [n]}$ are defined by uniformly choosing exactly $s$ of the variables per column to be a $1$). We present the details in Appendix \ref{sec:weakness}.
\section{Simple Proof of Theorem~$\ref{OurMainResult}$}
The main ingredient in our proof of Theorem~$\ref{OurMainResult}$ is the following bound on $\norm{Z}_{p}$:
\begin{lemma}
\label{mybound}
Let $B = m/s^2$. If $p \ge 2$, then
\[\norm{Z}_{p} \lesssim 
\begin{cases}
\frac{p}{s\log B}, & \text{ if } B \ge e \\
\frac{p}{sB} & \text{ if } B < e.
\end{cases}\]
\end{lemma}
\noindent We will later show that Theorem~$\ref{OurMainResult}$ follows from Lemma~$\ref{mybound}$ via Markov's inequality.

In order to analyze $\norm{Z}_p$, we view $Z$ as a quadratic form $\frac{1}{s}\sigma^T A \sigma$, where the vector $\sigma$ is an $n$-dimensional vector of independent Rademachers, and $A = (a_{i,j})$ is a symmetric, zero-diagonal $n \times n$ matrix where the $(i,j)$th entry (for $i \neq j$) is $x_i x_j \sum_{r=1}^m \eta_{r,i} \eta_{r,j}$. Since $Z$ is symmetric in $x_1, \ldots, x_n$, we can assume WLOG that $|x_1| \ge |x_2| \ge \ldots \ge |x_n|$. For convenience, we define, like in \cite{NelsonNotes},
\begin{equation}
\label{qdef}
Q_{i,j} := \sum_{r=1}^m \eta_{r,i} \eta_{r,j}
\end{equation}
to be the number of collisions between the nonzero entries of the $i$th column and the nonzero entries of the $j$th column. Now, the $(i,j)$th entry of $A$ (for $i \neq j$) can be written as $Q_{i,j}x_ix_j$. 

As discussed in Section 3.2, we cannot apply the Hanson-Wright bound to tightly analyze $\norm{Z}_p$ and thus require a separate treatment of Rademachers. We derive the following moment bound on quadratic forms of Rademachers\footnote{As mentioned before, Lata{\l}a \cite{LatalaChaos} provides a tight bound on the moments of $\sigma^T A \sigma$ (and on the moments of more general quadratic forms). However, his bound consists of terms that are difficult to analyze when the quadratic form coefficients are random variables. Moreover, his proof is quite complicated, though the bound can be used in a black box to generate a much messier solution (by unravelling some of his proof to avoid the operator-norm-like term).} that yields tighter guarantees than the Hanson-Wright bound yields for $\norm{Z}_p$:
\begin{lemma}
\label{MainQuadFormBound}
If $A= (a_{i,j})$ is a symmetric square $n \times n$ matrix with zero diagonal, $\left\{\sigma_i \right\}_{i \in [n]}$ is a set of independent Rademachers, and $q \ge 1$, then
\[\norm{\sum_{i=1}^n \sum_{j=1}^n a_{i,j}\sigma_i\sigma_j}_{q} \lesssim \left(\sum_{i=1}^{\min(q, n)} \sum_{j=1}^{\min(q,n)} |a_{i,j}|\right) + \sqrt{q} \sqrt{\sum_{i=1}^n \norm{\sum_{j > q} a_{i,j} \sigma_j}_{q}^2}.\]
\end{lemma}

Observe that our bound avoids the weakness of the Hanson-Wright bound in the limit as $q \rightarrow \infty$. As discussed in Section 1.2, $\norm{\sum_{i=1}^n \sum_{j=1}^n a_{i,j}\sigma_i\sigma_j}_{q}$ can be bounded by the entrywise $\ell_1$-norm bound $\sum_{i=1}^n \sum_{j=1}^n |a_{i,j}|$ for any $q \ge 1$. While the Hanson-Wright bound goes to $\infty$ as $q \rightarrow \infty$, the bound in Lemma~$\ref{MainQuadFormBound}$ approaches the entrywise $\ell_1$ bound in the limit: for $q > n$, the second term in Lemma~$\ref{MainQuadFormBound}$ vanishes since the summand $\sum_{j > q}$ is empty. As a result, the bound becomes the first-term, which becomes $\sum_{i=1}^n \sum_{j=1}^n |a_{i,j}|$ as desired.
For $1 \le q < n$, our bound becomes an interpolation of $\ell_1$ and $\ell_2$ norm bounds that bears resemblance to Hitczenko's Rademacher linear form bound in \cite{OriginalNorm} discussed in Section 1.2.

Although our bound is weaker than Lata{\l}a's bound in \cite{LatalaChaos} in the general case, it is much simpler to analyze, especially when $A$ is a random matrix. While the bound in \cite{LatalaChaos} is focused on obtaining tight estimates for quadratic forms where $A$ is a scalar matrix, our bound is much more tractable when $A$ is a random matrix. The main complication in the bound in \cite{LatalaChaos} arises from the operator-norm-like term $\sup_{||y||_2 = 1, ||y||_{\infty} \le \frac{1}{\sqrt{q}}} |y^TAy|$. Due to the asymmetrical geometry of the $\ell_2$ ball truncated by $\ell_{\infty}$ planes, this term becomes especially messy in our setting when $A$ is a random matrix. Observe that our bound in Lemma~$\ref{MainQuadFormBound}$ manages to avoid this term altogether. Moreover, our $\ell_1$ norm term is straightforward to calculate, and our $\ell_2$ norm term can be handled cleanly through a bound (Lemma~$\ref{sumsymm}$) from \cite{LatalaMoments} on the $q$-norm $\norm{\sum_{j > q} a_{i,j} \sigma_j}_{q}$ that is tractable even when the $a_{i,j}$ are themselves random variables.

We defer our proof of Lemma~$\ref{MainQuadFormBound}$ to Section 4.1. We now use Lemma~$\ref{MainQuadFormBound}$ and the triangle inequality to obtain the following bound on $||Z||_p$:
\begin{align*}
    ||Z||_p &= \frac{1}{s} \left(\mathbb{E}_{\eta} \mathbb{E}_{\sigma} \left[\sum_{i=1}^n \sum_{j \le n, j \neq i} Q_{i,j} x_i x_j \sigma_i\sigma_j \right]^p\right)^{1/p}
    \\
    &\lesssim \frac{1}{s} \left(\mathbb{E}_{\eta} \left[\sum_{i=1}^p \sum_{\substack{j \le p \\ j\neq i}} |Q_{i,j} x_i x_j| + \sqrt{p} \sqrt{\sum_{i=1}^n \left(\mathbb{E}_{\sigma} \left[\sum_{\substack{j > p \\ j\neq i}} Q_{i,j} x_i x_j \sigma_j\right]^p\right)^{2/p}} \right]^p\right)^{1/p}\\
    &\le \frac{1}{s} \left( \underbrace{\norm{\sum_{i=1}^p
    \sum_{\substack{j \le p \\ j\neq i}} |Q_{i,j} x_i x_j|}_{p}}_{(\ast)} + \underbrace{\sqrt{p} \sqrt{\sum_{i=1}^n \norm{\sum_{\substack{j > p \\ j\neq i}} Q_{i,j} x_i x_j \sigma_j}_{p}^2}}_{(\ast \ast)}\right).
\end{align*}

We first discuss some intuition for why using this bound to analyze $\norm{Z}_p$ avoids the loss incurred by the Hanson-Wright bound here. In the Hanson-Wright bound, all of the Rademachers are essentially approximated by gaussians. In our bound, we make use of Rademachers in the appropriate places to avoid loss. For $1 \le i \le p$ and $1 \le j \le p$ (the upper left $p \times p$ minor where the $|x_i|$ and $|x_j|$ values are the largest), our approach utilizes an $\ell_1$-norm bound rather than $\sqrt{p}$ times an $\ell_2$ bound, which turns out to allow us to save a factor of $\sqrt{p}$ in the resulting bound on $\norm{Z}_p$. Now, since the original matrix is symmetric, it only remains to consider $1 \le i \le n$ and $p+1 \le j \le n$. In this range, we approximate the $\sigma_i$ Rademachers by gaussians and use an $\ell_2$-norm bound. It turns out that approximating the $\sigma_j$ Rademachers by gaussians as well would yield too loose of a bound for our application, so we preserve the $\sigma_j$ Rademachers. For the remaining Rademacher linear forms, the interaction between the $x_j$ values (all of which are upper bounded in magnitude by $\frac{1}{\sqrt{p}}$) and the $\sigma_j$ Rademachers yields the desired bound. 

In order to prove Lemma~$\ref{mybound}$, it remains to prove Lemma~$\ref{MainQuadFormBound}$ as well as to bound $(\ast)$ and $(\ast \ast)$. In Section 4.1, we prove Lemma~$\ref{MainQuadFormBound}$. In Section 4.2 and Section 4.3, we bound $(\ast)$ and $(\ast \ast)$. Since the building blocks of $(\ast)$ and $(\ast \ast)$ are weighted sums of the $Q_{i,j}$ random variables, we first bound moments of these random variables separately. In Section 4.2, we use the binomial-like properties of the $Q_{i,j}$s coupled with standard moment bounds involving the binomial distribution to analyze the moments. In Section 4.3, we use these moment bounds to bound $(\ast)$ and $(\ast \ast)$, and then finish our proof of Lemma~$\ref{mybound}$. In Section 4.4, we show how Lemma~$\ref{mybound}$ implies Theorem~$\ref{OurMainResult}$.
\subsection{Proof of Lemma~$\ref{MainQuadFormBound}$}
We use the following standard lemmas in our proof of Lemma~$\ref{MainQuadFormBound}$.

The first lemma allows us to decouple the two sets of Rademachers in our quadratic form so that we can reduce analyzing the moments of the quadratic form to analyzing the moments of a linear form.  
\begin{lemma}[Decoupling, Theorem 6.1.1 of \cite{decouple}]
\label{decoupling}
If  $A=(a_{i,j})$ is a symmetric, zero-diagonal $n \times n$ matrix and $\left\{\sigma_i \right\}_{i \in [n]} \cup \left\{\sigma'_i \right\}_{i \in [n]}$ are independent Rademachers, then
\[\norm{\sum_{i=1}^n \sum_{j=1}^n a_{i,j}\sigma_i\sigma_j}_{q} \lesssim \norm{\sum_{i=1}^n \sum_{j=1}^n a_{i,j}\sigma'_i\sigma_j}_{q}.\]
\end{lemma}
The next lemma is due to Khintchine and gives an $\ell_2$-norm bound on linear forms of Rademachers. Since the Khintchine bound is derived from approximating $\sigma_1, \ldots, \sigma_n$ by i.i.d gaussians, we only use this bound outside of the upper left $p \times p$ minor of our matrix $A$.
\begin{lemma}[Khintchine]
\label{khint}
If $\sigma_1, \sigma_2, \ldots, \sigma_n$ are independent Rademachers, then for all $q \ge 1$ and $a \in \mathbb{R}^n$, 
\[\norm{\sum_{i=1}^n \sigma_i a_i}_{q} \lesssim \sqrt{q} ||a||_2. \]
\end{lemma} 

Now, we are ready to prove Lemma~$\ref{MainQuadFormBound}$. 
\begin{proof}[Proof of Lemma~$\ref{MainQuadFormBound}$]
By Lemma~$\ref{decoupling}$ and the triangle inequality, we know
\[\norm{\sum_{i=1}^n \sum_{j=1}^n a_{i,j}\sigma_i\sigma_j}_{q} \lesssim \underbrace{\norm{\sum_{i=1}^{\min(q,n)} \sum_{j=1}^{\min(q,n)} a_{i,j}\sigma'_i\sigma_j}_{q}}_{\alpha} + \underbrace{\norm{\sum_{i=1}^n \sum_{j>q} a_{i,j}\sigma'_i\sigma_j}_{q}}_{\beta} + \underbrace{\norm{\sum_{i>q} \sum_{j=1}^q a_{i,j}\sigma'_i\sigma_j}_{q}}_{\gamma}.\]
We first bound $\alpha$. Since a Rademacher $\sigma$ satisfies $|\sigma|=1$, it follows that $\alpha$ can be upper bounded by the entrywise $\ell_1$-norm bound $\sum_{i=1}^{\min(q,n)} \sum_{j=1}^{\min(q, n)} |a_{i,j}|$ as desired. Using Lemma~$\ref{khint}$, we know that $\beta$ can be upper bounded by:
\[\sqrt{q} \norm{\sqrt{\sum_{i=1}^n\left(\sum_{j>q} a_{i,j} \sigma_j\right)^2}}_{q} = \sqrt{q} \sqrt{\norm{\sum_{i=1}^n\left(\sum_{j>q} a_{i,j} \sigma_j\right)^2}_{q/2}} \le \sqrt{q} \sqrt{\sum_{i=1}^n\norm{\sum_{j>q} a_{i,j} \sigma_j}_{q}^2}.\]
We now bound $\gamma$. An analogous argument shows $\gamma \le \sqrt{q}\sqrt{\sum_{j=1}^q \norm{\sum_{i > q} a_{i,j} \sigma_i}_{q}^2}$. Thus: 
\[\gamma \le \sqrt{q}\sqrt{\sum_{j=1}^q \norm{\sum_{i > q} a_{i,j} \sigma_i}_{q}^2} \le \sqrt{q}\sqrt{\sum_{j=1}^n \norm{\sum_{i > q} a_{i,j} \sigma_i}_{q}^2} = \sqrt{q}\sqrt{\sum_{i=1}^n \norm{\sum_{j > q} a_{i,j} \sigma_j}_{q}^2}.\]
\end{proof}
\subsection{Moments of Weighted Sums of $Q_{i,j}$ Random Variables}
Recall that for $1 \le i \neq j \le n$, the $Q_{i,j}$ random variables count the number of collisions between the nonzero entries in the $i$th column and $j$th column. We first prove that these sets of random variables satisfy (conditional) independence properties, when conditioned on any choice of nonzero entries in the $i$th column. We also show that the moments of the random variables obtained through this conditioning are bounded by binomial moments.
\begin{proposition}
\label{QRV}
Let $X$ be a random variable distributed as $\Bin(s, s/m)$. For any $1 \le i \le n$, given any choice of $s$ nonzero rows $r_1 \neq r_2 \neq \ldots \neq r_s$ in the $i$th column, the set of $n-1$ random variables\footnote{\label{pspacechange}See Appendix \ref{appendix:probspace} for a formal discussion of viewing these quantities as random variables over a different probability space.} $\left\{(Q_{i,j} \mid \eta_{r_1, i} = \eta_{r_2, i} = \ldots = \eta_{r_s, i} = 1)\right\}_{1 \le j \le n, j \neq i}$ are independent. Moreover, for any $q \ge 1$ and any $j \neq i$:
\[\norm{Q_{i,j} \mid \eta_{r_1, i} = \eta_{r_2, i} = \ldots = \eta_{r_s, i} = 1}_{q} \le \norm{X}_{q}.\]
\end{proposition}
The independence properties use that the nonzero entries in different columns are independent. Moreover, the binomial bound on the moments of $Q_{i,j}$ follows from the decomposition of $Q_{i,j} \mid \eta_{r_1, i} = \eta_{r_2, i} = \ldots = \eta_{r_s, i} = 1$ into a sum of Bernoulli random variables.
\begin{proof}[Proof of Proposition~$\ref{QRV}$]
Let $A$ be a matrix drawn from $\mathcal{A}$, and pick any $1 \le i \le n$. We condition on the event that the $s$ nonzero entries in column $i$ of $A$ occur at rows $r_1, \ldots, r_s$. For $1 \le j \le n$, $j \neq i$ and $1 \le k \le s$, let $Y_{k,j} = \eta_{r_k, j}$, so that $(Q_{i,j} \mid \eta_{r_1, i} = \eta_{r_2, i} = \ldots = \eta_{r_s, i}) = \sum_{k=1}^s Y_{k,j}$. Notice that the sets $\left\{Y_{k,j} \right\}_{k \in [s]}$ for $1 \le j \le n, j \neq i$ are independent from each other, which means random variables in the set $\left\{Q_{i,j} \mid \eta_{r_1, i} = \eta_{r_2, i} = \ldots = \eta_{r_s, i} = 1 \right\}_{1 \le j \le n, j \neq i}$ are independent. For $1 \le j \le n, j \neq i$, and $1 \le k \le s$, let $Z_{k,j}$ be distributed as i.i.d Bernoulli random variables with expectation $s/m$. Notice that for a fixed $j$, each $Y_{k,j}$ is distributed as $Z_{k,j}$ and the random variables $\left\{ Y_{k,j} \right\}_{1 \le k \le s}$ are negatively correlated (and nonnegative), which means 
\[\norm{Q_{i,j} \mid \eta_{r_1, i} = \eta_{r_2, i} = \ldots = \eta_{r_s, i} = 1}_q = \norm{\sum_{k=1}^s Y_{k,j}}_{q} \le \norm{\sum_{k=1}^s Z_{k,j}}_{q} = \norm{X}_{q}.\]
\end{proof}

Now, we need to analyze the moments of weighted sums of $Q_{i,j}$ random variables. Using the independence properties and the fact that the moments of the $Q_{i,j}$ are upper bounded by binomial moments as given in Proposition~$\ref{QRV}$, this boils down to studying the moments of weighted sums of binomial random variables. The main tools that we use in analyzing these moments are bounds on moments of sums of nonnegative random variables and sums of symmetric random variables due to Lata{\l}a \cite{LatalaMoments} that we state in Appendix \ref{sec:latalamomentbounds}.\footnote{The proofs of these bounds given in \cite{LatalaMoments} are not complicated; for the sake of being self-contained, we give sketches of these proofs in Appendix \ref{sec:latalamomentbounds}.} 

Our first estimate is an upper bound on the moments of binomial random variables, which also gives bounds on moments of the $Q_{i,j}$ by Proposition~$\ref{QRV}$. We defer the proof to Appendix \ref{sec:ProofManyBin}. 
\begin{proposition}
\label{ManyBin}
Suppose that $X$ is a random variable distributed as $\Bin(N, \alpha)$ for any $\alpha \in (0,1)$ and any integer $N \ge 1$. If $q \ge 1$ and $B = \frac{q}{\alpha \max(N, q)}$, then 
\[\norm{X}_{q} \lesssim
\begin{cases}
\frac{q}{\log B} & \text{ if } B \ge e \\
\frac{q}{B}  & \text{ if } B < e
\end{cases}.\]
\end{proposition} 

Our next estimate is essentially an upper bound on the moments of sums of binomial random variables weighted by Rademachers. We defer the proof to Appendix \ref{sec:proofManyWeightedBin}.
\begin{proposition}
\label{ManyWeightedBin}
Suppose that $q \ge 2$ is even and $y = [y_1, \ldots, y_M]$ is a vector that satisfies $||y||_2 \le 1$ and $||y||_{\infty} \le \frac{1}{\sqrt{q}}$. Let $X$ be a random variable distributed as $\Bin(N, \alpha)$ for some $\alpha \in (0,1)$ and some integer $N \ge 1$. Suppose that $Y_1, \ldots, Y_M$ are independent random variables that  satisfy $||Y_k||_l \le ||X||_l$ for $1 \le k \le M$ and for $l \ge 1$. Suppose that $\sigma_1, \ldots, \sigma_M$ are independent Rademachers, also independent of $\left\{Y_k\right\}_{k \in [M]}$. If $B = \frac{q}{\alpha \max(N, q)}$, then
\[\norm{\sum_{k=1}^M Y_{k} y_k \sigma_k}_{q}  \lesssim
\begin{cases}
\frac{\sqrt{q}}{\log B} & \text{ if } B \ge e \\
\frac{\sqrt{q}}{B}  & \text{ if } B < e
\end{cases}.\]
\end{proposition}

\subsection{Bounding ($\ast$) and ($\ast \ast$) to prove Lemma~$\ref{mybound}$}
We bound the quantities ($\ast$) and ($\ast \ast$) in the following sublemmas, which assume the notation used throughout the paper:
\begin{lemma}
\label{termone}
If $m/s^2 = B$, then
\[\norm{\sum_{i=1}^{p}  \sum_{j \le p, j \neq i} |Q_{i,j} x_j x_i|}_{p}
\lesssim 
\begin{cases}
\frac{p}{\log B} & \text{ if } B \ge e \\
\frac{p}{B}  & \text{ if } B < e
\end{cases}
.\]
\end{lemma}
\begin{lemma}
\label{termtwo}
If $m/s^2 = B$, then
\[\sqrt{p} \sqrt{\sum_{i=1}^n \norm{\sum_{j > p, j \neq i} Q_{i,j} x_ix_j \sigma_j}_{p}^2}
\lesssim 
\begin{cases}
\frac{p}{\log B} & \text{ if } B \ge e \\
\frac{p}{B}  & \text{ if } B < e
\end{cases}
.\]
\end{lemma}

We now use Proposition~$\ref{QRV}$ as well as the moment bound on binomial random variables from Proposition~$\ref{ManyBin}$ to prove Lemma~\ref{termone} and thus bound $(\ast)$.
\begin{proof}[Proof of Lemma~$\ref{termone}$]
We carefully use the triangle inequality to see\footnote{Naively applying the triangle inequality yields a suboptimal bound, so we require this more careful treatment.}:
\[\norm{\sum_{i=1}^{p}  \sum_{\substack{j \le p \\ j \neq i}} |Q_{i,j} x_j x_i|}_{p} \le 2 \norm{\sum_{i=1}^{p} \sum_{\substack{j \le p \\ j > i}}  Q_{i,j} |x_j| |x_i|}_{p} \lesssim  \norm{\sum_{i=1}^p x^2_i \sum_{\substack{j \le p \\ j > i}} Q_{i,j}}_{p} \lesssim \sum_{i=1}^p x^2_i \norm{ \sum_{\substack{j \le p \\ j > i}}  Q_{i,j}}_{p}. \]
Let $X \sim \Bin(s, s/m)$ and $Y \sim \Bin(sp, s/m)$. By Proposition~$\ref{QRV}$, for any $i$ and any $r_1 \neq r_2 \neq \ldots \neq r_s$, the random variables $\left\{Q_{i,j} \mid \eta_{r_1, i} = \ldots = \eta_{r_s, i} = 1\right\}_{j \neq i}$ are independent and $\norm{Q_{i,j} \mid \eta_{r_1, i} = \ldots = \eta_{r_s, i} = 1}_p \le \norm{X}_p$. It follows from taking $p$th powers of both sides that 
\[\norm{\left(\sum_{\substack{j \le p \\ j > i}} Q_{i,j}\right) \mid \eta_{r_1, i} = \ldots = \eta_{r_s, i} = 1}_{p} = \norm{\sum_{\substack{j \le p \\ j > i}} (Q_{i,j} \mid \eta_{r_1, i} = \ldots = \eta_{r_s, i} = 1)}_{p} \leq \norm{Y}_p.\] Now, Proposition~$\ref{ManyBin}$ gives us a bound on $\norm{Y}_p$, and the result follows from the law of total expectation.\footnote{See Appendix \ref{appendix:probspace} for a formal discussion of why a uniform bound on the conditional $p$-norm implies a bound on the $p$-norm here.}
\end{proof}

We now use Proposition~$\ref{QRV}$ as well as the moment bound on weighted sums of binomial random variables from Proposition~$\ref{ManyWeightedBin}$ to prove Lemma~\ref{termtwo} and thus bound $(\ast \ast)$.
\begin{proof}[Proof of Lemma~$\ref{termtwo}$]
Observe that 
\[\sqrt{p} \sqrt{\sum_{i=1}^n \norm{\sum_{\substack{j > p \\ j \neq i}} Q_{i,j} x_ix_j \sigma_j}_{p}^2} = \sqrt{p} \sqrt{\sum_{i=1}^n x_i^2 \norm{\sum_{\substack{j > p \\ j \neq i}} Q_{i,j} x_j \sigma_j}_{p}^2} \le \sqrt{p} \max_{1 \le i \le n} \norm{\sum_{\substack{j > p \\ j \neq i}} Q_{i,j} x_j \sigma_j}_{p}.\]
Let $X \sim \Bin(s, s/m)$ and $Y \sim \Bin(sp, s/m)$. By Proposition~$\ref{QRV}$, for any $i$ and any $r_1 \neq r_2 \neq \ldots \neq r_s$, the random variables $\left\{Q_{i,j} \mid \eta_{r_1, i} = \ldots = \eta_{r_s, i} = 1\right\}_{j \neq i}$ are independent and $\norm{Q_{i,j} \mid \eta_{r_1, i} = \ldots = \eta_{r_s, i} = 1}_p \le \norm{X}_p \le \norm{Y}_p$. Moreover, $|x_j| \le \frac{1}{\sqrt{p}}$ for $j > p$. Now, we consider $\norm{\sum_{j > p, j \neq i} Q_{i,j} x_j \sigma_j \mid \eta_{r_1, i} = \ldots = \eta_{r_s, i} = 1}_{p}$ which is equal to 
\[\norm{\sum_{j > p, j \neq i} (Q_{i,j} \mid \eta_{r_1, i} = \ldots = \eta_{r_s, i} = 1) (\sigma_j \mid \eta_{r_1, i} = \ldots = \eta_{r_s, i} = 1)  x_j}_{p}.\] Since each $(\sigma_j \mid \eta_{r_1, i} = \ldots = \eta_{r_s, i} = 1)$ is distributed as a Rademacher and since the set of $n-1$ random variables $\left\{\sigma_j \mid \eta_{r_1, i} = \ldots = \eta_{r_s, i} = 1\right\}_{j \neq i}$ are independent and also independent of  $\left\{Q_{i,j} \mid \eta_{r_1, i} = \ldots = \eta_{r_s, i} = 1\right\}_{j \neq i}$, we can apply Proposition~$\ref{ManyWeightedBin}$ to this expression and thus get a bound\footnote{Approximating the $\sigma_j$ by gaussians yields a suboptimal bound, so we require the bound given in Proposition~$\ref{ManyWeightedBin}$.} on the conditional $p$-norm $\norm{\sum_{j > p, j \neq i} Q_{i,j} x_j \sigma_j \mid \eta_{r_1, i} = \ldots = \eta_{r_s, i} = 1}_{p}$. Now, the result follows from the law of total expectation.
\end{proof}

We now show the bound on $||Z||_p$ follows from the bounds on $(\ast)$ and $(\ast \ast)$ in Lemmas~$\ref{termone}, \ref{termtwo}$.
\begin{proof}[Proof of Lemma~$\ref{mybound}$]
Applying Lemmas~$\ref{termone},\ref{termtwo}$ after the following simplification proves the lemma:
\[\norm{Z}_{p} \lesssim \frac{1}{s}\norm{\sum_{i=1}^p \sum_{j \le p, j \neq i} |Q_{i,j} x_i x_j|}_{p} + \frac{\sqrt{p}}{s} \sqrt{\sum_{i=1}^n \norm{\sum_{j > p, j \neq i} Q_{i,j} x_i x_j \sigma_j}_{p}^2}.\]
\end{proof}
\subsection{Proof of Theorem~$\ref{OurMainResult}$}
We show Lemma~$\ref{mybound}$ implies Theorem~$\ref{OurMainResult}$, completing the proof.
\begin{proof}[Proof of Theorem~$\ref{OurMainResult}$]
It suffices to show $\mathbb{P}_{\eta, \sigma} [|Z| > \epsilon] < \delta$. By Markov's inequality, we know
\[\mathbb{P}_{\eta, \sigma} [|Z| > \epsilon] = \mathbb{P}_{\eta, \sigma} [|Z|^p > \epsilon^p] < \epsilon^{-p} \mathbb{E}[|Z|^p] = \left(\frac{\norm{Z}_{p}}{\epsilon}\right)^p.\]
Suppose that $B \ge e$. Then by Lemma~$\ref{mybound}$, we know
\[\left(\frac{\norm{Z}_{p}}{\epsilon}\right)^p \le \left(\frac{Cp}{(\log B)s\epsilon}\right)^p.\] Thus, to upper bound this quantity by $\delta$, we can set $s = \Theta(\eps^{-1}p/\log B) = \Theta(\eps^{-1}\log_B(1/\delta))$ and $m = \Theta(Bs^2)$. We impose the additional constraint that $B \le \frac{1}{\delta}$ to guarantee that $s \ge 1$. This proves the desired result.\footnote{If we set $B < e$, if we use Lemma~$\ref{mybound}$, we know that in order to obtain an upper bound of $\delta$, we would have to set $s = \Theta(\eps^{-1}p/B) = \Theta(\eps^{-1}\log(1/\delta)/B)$ and $m = \Theta(\eps^{-1}\log^2(1/\delta)/B)$. This yields no better $s$ or $m$ values than those achieved when $B = e$.} 
\end{proof}
\bibliography{bib.bib}
\appendix
\section{Limitations of the Combinatorial Approach}
At first glance, it appears that the bound of $\mathbb{P}[|Z| > \epsilon] \le \left(\frac{Ct s}{\epsilon m} \right)^t$ of \cite{Original} in the proof of the main theorem (p. 9 of the arXiv version) implies the desired dimension-sparsity tradeoffs by setting $s = \epsilon^{-1} t$, $m = \frac{B \epsilon^{-1} ts}{2C}$, and $t = \log_B(1/\delta)$ (this $t$ value is equivalent to the $p$ value in this paper). However, this does not actually follow from the analysis in \cite{Original}: there is an assumption made in one of the lemmas, which is not stated explicitly in the statement of the lemma, that does not allow the parameters to be set in this way. The limiting factor is the lemma that states that 
\[s^t \mathbb{E}[Z^t] \le 2^{O(t)} t^t \left(\frac{s^2}{m}\right)^t.\] This is Lemma 3 in the conference version of \cite{Original}, and Lemma 4.3 in the arXiv version of \cite{Original}. Here, $Z$ is defined analogously as in section 2.1 of this paper. 

The proof of this lemma, given in Appendix A.3 in \cite{Original}, implicitly relies on the fact that $\frac{s^2}{m} \ge 1$, although this condition is not explicitly stated in the lemma statement. This assumption arises from the last line of the proof, where the sum $\sum_{w=1}^t \left(\frac{s^2}{m}\right)^w$ is upper bounded by $t \left(\frac{s^2}{m} \right)^t$. Following the end of the proof of Theorem 1 (the top of p. 9 of the arXiv version), this yields $\mathbb{P}[|Z| > \epsilon] \le \left(\frac{Ct s}{\epsilon m} \right)^t$. Now, suppose we instead set $m = Bs^2$ (where $B \le 1$ as required by the assumption). Then we obtain $\left(\frac{Ct s}{\epsilon m} \right)^t = \left(\frac{Ct}{\epsilon Bs} \right)^t$. Thus, we can set $s$ to be $C\epsilon^{-1} B^{-1} \log(1/\delta)$ and $m$ to be $C^2 \epsilon^{-2} \log^2(1/\delta) B^{-1}$. Since $B \le 1$, this is no better than the original theorem statement and thus yields no dimension-sparsity tradeoff. 

Now, suppose we instead let $\frac{s^2}{m} \le 1$. Then we can modify the proof of Lemma 4.3 to obtain the weaker upper bound of $\sum_{w=1}^t \left(\frac{s^2}{m}\right)^w$ by $t \frac{s^2}{m}$. Let $B = m/s^2$ where $B \ge 1$. In order to ensure that $m$ is polynomial in $\log(1/\delta)$, we assume that $B \le \delta$. In this case, mimicking the calculation at the end of the proof of Theorem 1, we obtain $\mathbb{P}[|Z| > \epsilon] \le \frac{1}{B} \left(\frac{Ct}{\epsilon s} \right)^t = \left(\frac{Ct}{\epsilon s B^{1/t}} \right)^t$. Thus, we can set $s = C \log(1/\delta) \epsilon^{-1} e^{-\log B/t}$. Observe that $0 \le \log B \le t$, so $1 \ge e^{-\log B/t} \ge e^{-1}$. Thus, $s = \Theta(\log(1/\delta) \epsilon^{-1})$ and $m = \Theta(Bs^2)$, which does not yield a dimension-sparsity tradeoff.

Thus, it is not clear how to directly obtain the dimension-sparsity tradeoff from the combinatorial approach of \cite{Original}. Some intuition for this limitation is that the moment bounds on $Z$ obtained by the combinatorial approach are not sufficiently tight for varying values of $B$ due to the fact that the bounding techniques are implicitly tailored to the case of $B = \Theta(1)$. The combinatorics-free approach in this paper avoids this issue through making use of a more structured method to bound the moments of $Z$. 

\section{Formal Definition of Probability Space and Random Variables}\label{appendix:probspace}

\subsection{Construction of the Probability Space}
Rather than defining our probability space over $m \times n$ sparse, sign-consistent matrices implicitly by random variables, we explicitly define this (finite) probability space here. (For all probability spaces defined, the set of events will be all subsets of the sample space.) We take our probability space to be the product space of $n$ probability spaces $(\Omega_i, \Sigma_i, \mathbb{P}_i)$ where $\Omega_i$ is the set of $m$-dimensional column vectors with entries in $\left\{-1 / \sqrt{s}, 0, 1 / \sqrt{s} \right\}$ with exactly $s$ nonzero entries and all nonzero entries the same sign. Thus, the product $\Omega := \Omega_1 \times \Omega_2 \times \ldots \times \Omega_n$ is in one-to-one correspondence with the sample space of $m \times n$ dimensional sign-consistent matrices with entries in  $\left\{-1 / \sqrt{s}, 0, 1 / \sqrt{s} \right\}$ with exactly $s$ nonzero entries in each column. Now, we define the probability measure $\mathbb{P}_i$ for each $1 \le i \le n$, which implicitly defines the product probability measure $\mathbb{P}$ using the definition of a product of probability spaces. 

We now define $(\Omega_i, \Sigma_i, \mathbb{P}_i)$ as a product of probability spaces $(\Omega_i^1, \Sigma_i^1, \mathbb{P}_i^1) \times (\Omega_i^2, \Sigma_i^2, \mathbb{P}_i^2)$ which will separate sign from the choice of nonzero entries. Here, $\Omega_i^1$ is the set $\left\{-1, 1\right\}$ and $\Omega_i^2$ is the set of binary $m$-dimensional column vectors with exactly $s$ nonzero entries. Notice there is a bijection $\Omega_i^1 \times \Omega_i^2 \rightarrow \Omega_i$ defined by $(\omega_i^1, \omega_i^2) \mapsto (\frac{1}{\sqrt{s}} \omega_i^1) \omega_i^2$. Now, we define $\mathbb{P}^1_i(-1) = \mathbb{P}^1_i(1) = 0.5$. We take $\mathbb{P}^2_i$ to be any probability measure that satisfies the following properties: We use the notation that for $\omega_i^2 \in \Omega_i^2$, $\omega_i^2(r)$ is the $r$th coordinate of $\omega_i^2$.  First, for all $1 \le r \le m$, it must be true that $\sum_{w_i^2 \in \Sigma_i^2 \text{ such that }\omega_i^2(r) = 1} \mathbb{P}^2_i[\omega_i^2] = \frac{s}{m}$. Second, 
 we assume that for any subset $S \subseteq [m]$:
\[\sum_{\omega_i^2 \in \Omega_i^2} \mathbb{P}_i^2[\omega_i^2] \left(\prod_{r \in S} \omega_i^2(r)\right) \le \prod_{r \in S} \left(\sum_{\omega_i^2 \in \Omega_i^2} \mathbb{P}_i^2[\omega_i^2] \omega_i^2(r) \right).\]
\noindent This defines $\mathbb{P}_i$ using the definition of a product of probability spaces and the bijection from $\Omega_i^1 \times \Omega_i^2 \rightarrow \Omega_i$ described above. 

Let $\sigma_i: \Omega \rightarrow \left\{-1, 1 \right\}$ be the random variable for the sign of the $i$th column. Let $\eta_{r,i}: \Omega \rightarrow \left\{0, 1 \right\}$ be the indicator random variable for whether the $(r,i)$th entry is nonzero. Observe that the independence and negative correlation properties are recovered exactly. 

\subsection{Explanation of Proposition~$\ref{QRV}$}
As before, for $i \neq j$, we define $Q_{i,j} = \sum_{r=1}^m \eta_{r,i} \eta_{r,j}$. For some $r_1 \neq r_2 \neq \ldots \neq r_s$, let $A_{r_1, \ldots, r_s} \subseteq \Omega$ be the event that the nonzero entries in the $i$th column occur at $r_1, \ldots, r_s$. Now, $(Q_{i,j} \mid A_{r_1, \ldots, r_s})$ is a random variable defined by the restriction of the function $Q_{i,j}: \Omega \rightarrow \mathbb{R}$ to a function $A_{r_1, \ldots, r_s} \rightarrow \mathbb{R}$ on the probability space $(A_{r_1, \ldots, r_s}, 2^{A_{r_1, \ldots, r_s}}, \mathbb{P}|_{A_{r_1, \ldots, r_s}})$ where for $B \subseteq A_{r_1, \ldots, r_s}$, we define $\mathbb{P}|_{A_{r_1, \ldots, r_s}}(B)$ to be $\frac{\mathbb{P}(B)}{\mathbb{P}(A_{r_1, \ldots, r_s})}$. This formally defines the random variable $(Q_{i,j} \mid \eta_{r_1, i} = \eta_{r_2, i} = \ldots = \eta_{r_s, i} = 1)$ in Proposition~$\ref{QRV}$, and observe that the proof proceeds by computing the distribution of this random variable. In this notation, the proposition shows that the random variables $(Q_{i,j} \mid A_{r_1, \ldots, r_s})$ are independent on this new probability space. 

\subsection{Explanation of Lemma~$\ref{termone}$}
For Lemma~$\ref{termone}$, the random variable $\left(\sum_{i < j \le p} Q_{i,j}\right) \mid A_{r_1, \ldots, r_s}$ is defined analogously on the restricted probability space. It follows from definition that this random variable is equivalent to the sum of random variables $\sum_{i < j \le p} (Q_{i,j} \mid A_{r_1, \ldots, r_s})$. Since we know that $(Q_{i,j} \mid A_{r_1, \ldots, r_s})$ for $i < j \le p$ are independent and have moments bounded by binomials by Proposition~$\ref{QRV}$, this implies that we can apply Proposition~$\ref{ManyBin}$ to $\sum_{i < j \le p} (Q_{i,j} \mid A_{r_1, \ldots, r_s})$ to obtain a bound on the $p$-norm $\mathbb{E}\left[\left(\sum_{i < j \le p} Q_{i,j} \mid A_{r_1, r_2, \ldots, r_s}\right)^p\right]^{1/p}$ which gives us a bound on the conditional $p$-norm $\mathbb{E}\left[\left(\sum_{i < j \le p} Q_{i,j}\right)^p \mid A_{r_1, r_2, \ldots, r_s}\right]^{1/p}$. Let's suppose this bound is $\alpha$. Since the events $\left\{A_{r_1, \ldots, r_s}\right\}_{r_1 \neq r_2 \neq \ldots \neq r_s}$ form a disjoint partition of the probability space, we know by the law of total expectation that:
\[\alpha^p \ge \sum_{r_1 \neq r_2 \neq \ldots \neq r_s} \mathbb{P}[A_{r_1, r_2, \ldots, r_s}] \mathbb{E}\left[\left(\sum_{i < j \le p} Q_{i,j}\right)^p \mid A_{r_1, r_2, \ldots, r_s}\right] = \mathbb{E}\left[\left(\sum_{i < j \le p} Q_{i,j}\right)^p\right], \]
which means that $\norm{\sum_{i < j \le p} Q_{i,j}}_p \le \alpha$, so a uniform bound on the conditional $p$-norm implies a bound on the $p$-norm, which finishes the proof.

\subsection{Explanation of Lemma~$\ref{termtwo}$}
For Lemma~$\ref{termtwo}$, the random variable $\sum_{j > p, j \neq i} Q_{i,j} \sigma_j \mid A_{r_1, \ldots, r_s}$ is defined analogously on the restricted probability space. It follows from definition that this random variable is equivalent to the sum of random variables $\sum_{j > p, j \neq i} (Q_{i,j} \sigma_j \mid A_{r_1, \ldots, r_s}) = \sum_{j > p, j \neq i} (Q_{i,j} \mid A_{r_1, \ldots, r_s}) (\sigma_j \mid A_{r_1, \ldots, r_s})$. By Proposition~$\ref{QRV}$, we know that the random variables $(Q_{i,j} \mid A_{r_1, \ldots, r_s})$ are independent for $j > p, j \neq i$ and have moments bounded by binomials. Moreover, from the construction of the probability space, we know that $(\sigma_j \mid A_{r_1, \ldots, r_s})$ are independent for $j > p, j \neq i$ and are independent of all $(Q_{i,j} \mid A_{r_1, \ldots, r_s})$, and we also know that $(\sigma_j \mid A_{r_1, \ldots, r_s})$ is distributed as a Rademacher. This implies that we can apply Proposition~$\ref{ManyWeightedBin}$ to obtain a bound on the $p$-norm $\mathbb{E}\left[\left|\sum_{j > p, j \neq i} (Q_{i,j} \mid A_{r_1, \ldots, r_s}) (\sigma_j \mid A_{r_1, \ldots, r_s})\right|^p\right]^{1/p}$ which we can view as a bound on the conditional $p$-norm $\mathbb{E}\left[\left|\sum_{i < j \le p} Q_{i,j} x_j \sigma_j \right|^p \mid A_{r_1, r_2, \ldots, r_s}\right]^{1/p}$. Let's suppose that this bound is $\alpha$. Since the events $\left\{A_{r_1, \ldots, r_s}\right\}_{r_1 \neq r_2 \neq \ldots \neq r_s}$ form a disjoint partition of the probability space, we know by the law of total expectation that:
\[\alpha^p \ge \sum_{r_1 \neq r_2 \neq \ldots \neq r_s} \mathbb{P}[A_{r_1, r_2, \ldots, r_s}] \mathbb{E}\left[\left|\sum_{i < j \le p} Q_{i,j} \sigma_j x_j \right|^p \mid A_{r_1, r_2, \ldots, r_s}\right] = \mathbb{E}\left[\left|\sum_{i < j \le p} Q_{i,j} \sigma_j x_j \right|^p\right], \]
which means that $\norm{\sum_{i < j \le p} Q_{i,j} \sigma_j x_j}_p \le \alpha$, so a uniform bound on the conditional $p$-norm implies a bound on the $p$-norm, which finishes the proof.


\section{Lata{\l}a's Moment Bounds and Proof Sketches}\label{sec:latalamomentbounds}
We sketch proofs of the upper bounds given in the following two lemmas, due to Lata{\l}a. Full proofs of these lemmas can be found in \cite{LatalaMoments}.
\begin{lemma}[\cite{LatalaMoments}]
\label{sumsymm}
If $q \ge 2$ and $X, X_1, \ldots, X_n$ are independent symmetric random variables, then
\[\norm{\sum_{i=1}^n X_i}_{q} \simeq \inf \left\{T > 0 \text{ such that } \sum_{i=1}^n \log \left(\mathbb{E} \left[\left(1 + \frac{X_i}{T}\right)^q \right] \right) \le q \right\}.\]
\end{lemma}
\begin{lemma}[\cite{LatalaMoments}\footnote{This result was actually first due to S.J. Montgomery-Smith through a private communication with Lata{\l}a. Nonetheless, it is also a corollary of a result in \cite{LatalaMoments}.}]
\label{estimate}
If $1 \le q \le n$ and $X, X_1, \ldots, X_n$ are i.i.d nonnegative random variables, then
\[\norm{\sum_{i=1}^n X_i}_{q} \simeq \sup_{1 \le t \le q} \frac{q}{t} \left(\frac{n}{q} \right)^{1/t}\norm{X}_{t}.\]
\end{lemma}

For a random variable $X$, we define
\[\phi_q(X) := \mathbb{E}[|1+X|^q].\]
We begin with the following proposition that relates $\phi_q$ to the $q$-norm, which is useful in proving Lemma~$\ref{estimate}$ and Lemma~$\ref{sumsymm}$.
\begin{proposition}
\label{MainLatProof}
If independent random variables $X_1, \ldots, X_n$ and value $q \ge 1$ satisfy the following inequality for any $T > 0$: 
\[\left(\norm{\sum_{i=1}^n \frac{X_i}{T}}_{q}\right)^q \le \prod_{i=1}^n \phi_q\left(\frac{X_i}{T}\right),\] then
\[\norm{\left(\sum_{i=1}^n X_i\right)}_{q} \lesssim \inf \left\{T > 0 \text{ such that }\sum_{i=1}^n \log \left( \mathbb{E}\left[\left(1+\frac{X_i}{T}\right)^q\right] \right) \le q \right\}.\]
\end{proposition}
\begin{proof}
Suppose that $\sum_{i=1}^n \log \left( \mathbb{E}\left[\left(1+\frac{X_i}{T}\right)^q\right] \right) \le q$. Then,
\[\norm{\left(\sum_{i=1}^n X_i\right)}_{q} = T \norm{\left(\sum_{i=1}^n \frac{X_i}{T}\right)}_{q} \le T \left(\prod_{i=1}^n \phi_q\left(\frac{X_i}{T}\right)\right)^{1/q} = T e^{\frac{1}{q} \log \left(\sum_{i=1}^n \phi_q\left(\frac{X_i}{T}\right)\right)} \lesssim T.\]
\end{proof}
The proofs of the upper bounds of Lemma~$\ref{sumsymm}$ and Lemma~$\ref{estimate}$ boil down to showing that the condition of Proposition~$\ref{MainLatProof}$ is satisfied. In Section C.1, we sketch a proof of the upper bound of Lemma~$\ref{estimate}$. In Section C.2, we sketch a proof of the upper bound of Lemma~$\ref{sumsymm}$. 
\subsection{Proof Sketch of Lemma~$\ref{sumsymm}$ (Upper Bound)}
We first state the following two propositions. The proof of these propositions are straightforward calculations and can be found in \cite{LatalaMoments}.
\begin{proposition}
\label{productboundsymm}
If $q \ge 2$ and $Y_1, Y_2, \ldots, Y_n$ are independent symmetric random variables, then
\[\phi_{q}\left(\sum_{i=1}^n Y_i\right) \le \prod_{i=1}^n \phi_q(Y_i).\]
\end{proposition}
\begin{proposition}
\label{normboundsymm}
If $q \ge 2$ and $Y$ is a symmetric random variable, then
\[||Y||_q^q \le \phi_q(Y).\]
\end{proposition}
From Proposition~$\ref{productboundsymm}$ and Proposition~$\ref{normboundsymm}$, coupled with the fact that $X_1/T, X_2/T, \ldots, X_n/T$, and $\sum_{i=1}^n X_i/T$ are independent symmetric random variables, it follows that the condition of Proposition~$\ref{MainLatProof}$ is satisfied. 
\subsection{Proof Sketch of Lemma~$\ref{estimate}$ (Upper Bound)}
We first sketch the proof of the upper bound of the following sublemma, which is analogous to Lemma~$\ref{sumsymm}$:
\begin{lemma}[Lata{\l}a\cite{LatalaMoments}]
\label{estimategen}
If $q \ge 1$ and $X, X_1, \ldots, X_n$ are independent nonnegative random variables then
\[\norm{\sum_{i=1}^n X_i}_{q} \simeq \inf \left\{T > 0 \text{ such that } \sum_{i=1}^n \log \left(\mathbb{E} \left[\left|1 + \frac{X_i}{T}\right|^q \right] \right) \le q \right\}.\]
\end{lemma}
\noindent The upper bound of Lemma~$\ref{estimategen}$ follows from the following propositions, coupled with the fact that $X_1/T$, $X_2/T$, \ldots, $X_n/T$, and $\sum_{i=1}^n X_i/T$ are independent nonnegative random variables. The proofs of these propositions are straightforward calculations and can be found in \cite{LatalaMoments}.
\begin{proposition}
If $q \ge 1$ and $Y_1, Y_2, \ldots, Y_n$ are independent nonnegative random variables, then 
\[\phi_{q}\left(\sum_{i=1}^n X_i\right) \le \prod_{i=1}^n \phi_q(X_i).\]
\end{proposition}
\begin{proposition}
If $q \ge 1$ and $Y$ is a nonnegative random variable, then
\[ \norm{Y}_{q}^q \le \phi_q(Y).\]
\end{proposition}

Now, we describe how the upper bound of Lemma~$\ref{estimategen}$ implies the upper bound of Lemma~$\ref{estimate}$. It suffices to show if we take 
\[T = 2e\sup_{1 \le t \le q} \frac{q}{t} \left(\frac{n}{q} \right)^{1/t} \norm{X}_{t},\] then 
\begin{equation}
\label{desiredcond}
\sum_{i=1}^n \log \left(\mathbb{E} \left[\left|1 + \frac{X_i}{T}\right|^q \right] \right) \le q.
\end{equation} 
Since $q \le n$, it follows that
\begin{align*}
\mathbb{E} \left[\left|1 + \frac{X_i}{T}\right|^q \right] &= \frac{\mathbb{E}[X_i^q]}{T^q} + \sum_{k=0}^{q-1} {q \choose k} \frac{\mathbb{E}[X_i^k]}{T^k} \\
&\le \frac{\mathbb{E}[X_i^q]}{\left(\left(\frac{n}{q}\right)^{1/q}\norm{X_i}_q\right)^q} + \sum_{k=0}^{q-1} \left(\frac{qe}{k} \right)^k \frac{\mathbb{E}[X_i^k]}{\left(\frac{2eq\norm{X_i}_k}{k} \right)^k} \\
&\le \frac{q}{n} + 1 \\
&\le e^{\frac{q}{n}}
\end{align*}
from which \eqref{desiredcond} follows.

\section{Proof of Proposition~$\ref{ManyWeightedBin}$}\label{sec:proofManyWeightedBin}
The main ingredient in this proof is Lemma~$\ref{sumsymm}$ (Lata{\l}a'a bound on moments of sums of symmetric random variables). 
\begin{proof}[Proof of Proposition~$\ref{ManyWeightedBin}$]
Since the $Y_i$ are independent random variables, we can apply Lemma~$\ref{sumsymm}$ to obtain:
\[\norm{\sum_{k=1}^M Y_k y_k \sigma_k}_{q} \lesssim \inf \left\{ T> 0 \text{ } \rvert \sum_{k=1}^M \log \left(\mathbb{E} \left[\left|1 + \frac{Y_k\sigma_k y_k }{T}\right|^q \right] \right) \le q\right\}. \]
Thus, it suffices to show 
\[T \simeq
\begin{cases}
\frac{\sqrt{q}}{\log B} &\text{ if } B \ge e \\
\frac{\sqrt{q}}{B} &\text{ if } B < e \\
\end{cases}\] satisfies
 $\sum_{k=1}^M \log \left(\mathbb{E} \left[\left(1 + \frac{Y_k\sigma_k y_k }{t}\right)^q \right] \right) \le q$. We see
\begin{align*}
\sum_{k=1}^M \log \left(\mathbb{E} \left[\left(1 + \frac{Y_k\sigma_k y_k }{T}\right)^q \right] \right)
&=\sum_{k=1}^M  \log \left(1 + \sum_{l=1}^q {q \choose l} \frac{(\mathbb{E}[(Y_k\sigma_k)^l])y_k^l}{T^l}  \right)\\
&= \sum_{k=1}^M  \log \left(1 + \sum_{l=1}^{q/2} {q \choose {2l}} \frac{\norm{Y_k}^{2l}_{2l}y_k^{2l}}{T^{2l}}  \right) \\
&\le \sum_{k=1}^M \log \left(1 + \sum_{l=1}^{q/2} \left(\frac{qe}{2l}\right)^{2l} \left(\frac{\norm{Y_k}_{2l}y_k}{T}  \right)^{2l} \right)
\end{align*}
By the bound on moments of binomial random variables in Proposition~$\ref{ManyBin}$, we know if $B \ge e$ that there exists a universal constant $C$ such that $||Q_{i,j}||_{2l} \le \frac{2lC}{\log B}$. Thus, we obtain
\begin{align*}
\sum_{k=1}^M \log \left(\mathbb{E} \left[\left(1 + \frac{Y_k\sigma_k y_k}{T}\right)^q \right] \right) &\le\sum_{k=1}^M \log \left(1 + \sum_{l=1}^{q/2} \left(\frac{qe}{2l}\right)^{2l} \left(\frac{2lCy_k}{T\log B}  \right)^{2l} \right) \\
&\le \sum_{k=1}^M  \log \left(1 + \sum_{l=1}^{q/2} \left(\frac{qeCy_k}{T \log B}\right)^{2l}\right). \\
\end{align*}
Since $|y_k| \le \frac{1}{\sqrt{q}}$, if we set $T = \frac{2eC\sqrt{q}}{\log B}$, then we obtain
\[\sum_{k=1}^M  \log \left(1 + \sum_{l=1}^{q/2} \left(\frac{\sqrt{q}y_k}{2}\right)^{2l}\right) \le \sum_{k=1}^M  \log \left(1 + (\sqrt{q} y_k)^2 \sum_{l=1}^{q/2}  \left(\frac{1}{2}\right)^{2l}\right).\]
This can be bounded by
\[\sum_{k=1}^M \log \left(1 + \left(\sqrt{q}y_k \right)^{2}  \right) = \sum_{k=1}^M \log \left(1 + qy_k^2 \right) \le \sum_{i=1}^n qy_k^2 \le q\] as desired. An analogous argument shows that if $B < e$, we can set $T = \frac{2eC\sqrt{q}}{B}$.

\end{proof}

\section{Proof of Proposition~$\ref{ManyBin}$}\label{sec:ProofManyBin}
The main tool that we use in this proof is Lemma~$\ref{estimate}$ (Lata{\l}a's bound on moments of sums of i.i.d nonnegative random variables). 
\begin{proof}[Proof of Proposition~$\ref{ManyBin}$]
Notice that it suffices to obtain an upper bound on $\norm{X}_q$ for all $N \ge q$. (Since $\norm{X}_q$ is an increasing function of $N$, an upper bound on $\norm{X}_q$ at $N = q$ is also an upper bound on $\norm{X}_q$ for all $N < q$). For the rest of the proof, we assume $N \ge q$. 

Notice $X$ has the same distribution as $\sum_{j=1}^{N} Z_j$ where $Z, Z_1, \ldots, Z_{N}$ are i.i.d Bernoulli random variables with expectation $\alpha$. Since $\norm{Z}_t = \alpha^{1/t}$, we know by Lemma~$\ref{estimate}$,
\begin{align*}
\norm{X}_{q} &\simeq \sup_{1 \le t \le q} \frac{q}{t} \left(\frac{N}{q} \right)^{1/t} \alpha^{1/t} \\
&= \sup_{1 \le t \le q} \frac{q}{t} \left(\frac{1}{B} \right)^{1/t}
\end{align*}
At $t = 1$, this quantity is equal to $\frac{q}{B}$, and at $t=q$, this quantity is equal to $\left(\frac{1}{B} \right)^{1/q} = e^{\log (1/B)/ q}$. The only $t \in \mathbb{R}$ for which this quantity has derivative $0$ is $t = \log B$. Notice that $1 \le \log B \le q$ if and only if $e \le B \le e^q$. Thus
\[\norm{X}_q \simeq
\begin{cases}
\max(\frac{q}{B}, \frac{q}{\log B}, e^{\log(1/B)/q}) & \text{ if } e \le B \le e^q\\
\max(\frac{q}{B}, e^{\log(1/B)/q}) & \text{ if } B  < e \text{ or if } B > e^q.
\end{cases}.\]

For $B \ge e$, we want to show $\norm{X}_{q} \lesssim q/\log B$. Since $\log B > 0$, we see $e^{\log(1/B)/q} = e^{-\log B/q} \le q/\log B$ and $q/B \le q/\log B$. 

For $B < e$, we want to show $\norm{X}_{q} \lesssim q/B$. Since $\frac{1}{B} > \frac{1}{e}$, we see $e^{\log (1/B)/q} = \left(\frac{1}{B} \right)^{1/q} \le \frac{e}{B} \lesssim \frac{q}{B}$. 
\end{proof}
\section{Weakness of bound on $\norm{Z}_{p}$ from Equation $\eqref{failHW}$}\label{sec:weakness}
Like in Section 4, we view the random variable $Z$ as a quadratic form $\frac{1}{s}\sigma^TA\sigma$, where $\sigma$ an $n$-dimensional vector of independent Rademachers and $A$ is a symmetric, zero-diagonal $n \times n$ matrix where the $(i,j)$th entry (for $i \neq j$) is $x_ix_j\sum_{r=1}^m \eta_{r,i}\eta_{r,j} = Q_{i,j}x_ix_j$. Applying the Hanson-Wright bound followed by an expectation over the $\eta$ values yields
\begin{equation}
\label{failHW}
\norm{\sigma^TA\sigma}_{p} \lesssim 
\norm{\sqrt{p} \sqrt{\sum_{i=1}^{n} \sum_{j \le n, j \neq i} Q^2_{i,j}x_i^2 x_j^2} + p \sup_{\norm{y}_2 = 1} \left|\sum_{i=1}^n \sum_{j \le n, j \neq i} Q_{i,j}x_ix_jy_iy_j \right|}_{p} =: U_p.
\end{equation}

We show that the vector $x = [\frac{1}{\sqrt{2}}, \frac{1}{\sqrt{2}}, 0, \ldots, 0] \in \mathbb{R}^n$ forces $U_p$ to be too large to yield the optimal $m$ value, thus proving that the Hanson-Wright bound does not provide a sufficiently tight bound on $\norm{Z}_p$ to achieve Theorem~$\ref{OriginalMainResult}$. The main ingredient in our proof is the following lemma, which we prove in subsection C.1:
\begin{lemma}
\label{HWFailbound}
For every column $1 \le i \le n$, suppose that the random variables $\left\{\eta_{r,i}\right\}_{r \in [m], i \in [n]}$ have the distribution defined by uniformly choosing exactly $s$ of the variables per column. If $x = \left[\frac{1}{\sqrt{2}}, \frac{1}{\sqrt{2}}, 0, \ldots, 0 \right]$, $p < s$ and $B = m/s^2 \le \frac{e^p}{p}$, then
\[
U_p \simeq 
\begin{cases}
\frac{p^2}{\log Bp} & \text{ if } B \ge \frac{e}{p} \\
\frac{p}{B} & \text{ if } B < \frac{e}{p}.
\end{cases}
\]
\end{lemma}

We can obtain bounds on $s$ and $m$ from Lemma~$\ref{HWFailbound}$ via Markov's inequality. We disregard the case where $B \ge \frac{e^p}{p}$, since this case would yield a value for $m$ that is not polynomial in $\log(1/\delta)$. If $B < e/p$, then it follows that $s = \Theta(\eps^{-1} B^{-1} \log(1/\delta)) = \Omega(\eps^{-1} \log^2(1/\delta))$ and $m = \Theta(\eps^{-2}B^{-1} \log^2(1/\delta)) = \Omega(\eps^{-2} \log^3(1/\delta))$. If $B \ge e/p$, then it follows that $s = \Theta(\eps^{-1}p^2/\log(Bp)) = \Omega(\eps^{-1}\log(1/\delta))$ and $m = \Theta(\eps^{-2}p^4B/\log^2(Bp)) = \Omega(\eps^{-2}\log^3(1/\delta))$. These bounds on $m$ incur an extra $\log (1/\delta)$ factor, and thus the Hanson-Wright bound is too weak for this setting. Now, it suffices to prove Lemma~$\ref{HWFailbound}$, which we do in the next section.
\subsection{Proof of Lemma~$\ref{HWFailbound}$}
In this section, we assume that $x = \left[\frac{1}{\sqrt{2}}, \frac{1}{\sqrt{2}}, 0, \ldots, 0 \right]$ and that the random variables $\left\{\eta_{r,i}\right\}_{r \in [m], i \in [n]}$ have the distribution defined by uniformly choosing exactly $s$ of the variables per column. We first show the following computation of $||Q_{i,j}||_p$.
\begin{proposition}
\label{eqasym}
Assume that the random variables $\left\{\eta_{r,i}\right\}_{r \in [m], i \in [n]}$ have the distribution defined by uniformly choosing exactly $s$ of the variables per column. Then, if $p < s$ and $X \sim \Bin(s, s/m)$, we have that $||Q_{i,j}||_p \simeq ||X||_p$.
\end{proposition}
\begin{proof}
We condition on the even that the nonzero locations in column $i$ are at $r_1, r_2, \ldots, r_s$. Notice that the random variable $(Q_{i,j} \mid \eta_{r_1, i} = \eta_{r_2, i} = \ldots = \eta_{r_s, i} = 1)$ is distributed as $Z_{r_1} + Z_{r_2} + \ldots + Z_{r_s}$ where $Z_{r_k}$ is an indicator for the $k$th entry in the $j$th column being nonzero. Let $Z'_{r_k}$ for $1 \le k \le s$ be i.i.d random variables distributed as $\text{Bern}(s/m)$. Now, observe that \[\mathbb{E}[(Z_{r_1} + Z_{r_2} + \ldots + Z_{r_s})^p] = \sum_{\substack{0 \le t_1, t_2, \ldots, t_s \le p \\ t_1 + t_2 + \ldots + t_s = p}}\mathbb{E}[ \prod_{i=1}^s Z_{r_i}^{t_i}] =\sum_{\substack{0 \le t_1, t_2, \ldots, t_s \le p \\ t_1 + t_2 + \ldots + t_s = p}} \mathbb{E}[ \prod_{i \mid t_i > 0} Z_{r_i}].\] 

Notice that $\mathbb{E}[(Z'_{r_1} + Z'_{r_2} + \ldots + Z'_{r_s})^p] = \sum_{0 \le t_1, t_2, \ldots, t_s \le p, t_1 + t_2 + \ldots + t_s = p} \mathbb{E}[ \prod_{i \mid t_i > 0} Z'_{r_i}]$. Thus, it suffices to compare $\mathbb{E}[ \prod_{i \mid t_i > 0} Z_{r_i}]$ and $\mathbb{E}[ \prod_{i \mid t_i > 0} Z'_{r_i}]$. We see that $\mathbb{E}[ \prod_{i \mid t_i > 0} Z'_{r_i}] = \left(\frac{s}{m} \right)^{| \left\{i \mid t_i > 0 \right\}|}$. Since $p < s$, we see that $\mathbb{E}[ \prod_{i \mid t_i > 0} Z_{r_i}] = \prod_{j=0}^{|\left\{i \mid t_i > 0 \right\}| - 1} \frac{s-j}{m-j}$. It is not difficult to verify that this ratio is bounded by $2^{O(p)}$ as desired, so 
\[\frac{\mathbb{E}[\left(Q_{i,j} \mid \eta_{r_1, i} = \eta_{r_2, i} = \ldots = \eta_{r_s, i} = 1\right)^p]}{\mathbb{E}[X^p]} = \frac{\mathbb{E}[(Z_{r_1} + Z_{r_2} + \ldots + Z_{r_s})^p]}{\mathbb{E}[X^p]} \ge 2^{-O(p)}.\] Now, by the law of total expectation, we know that
\[\frac{\mathbb{E}[Q_{i,j}^p]}{\mathbb{E}[X^p]} \ge 2^{-O(p)}\]
as desired.
\end{proof}
We now prove the following relation between $U_p$ and $\norm{Q_{1,2}}_p$: 
\begin{lemma}
\label{keyFail}
Assume the notation and restrictions above. Then $U_p \simeq p \norm{Q_{1,2}}_{p}$. 
\end{lemma}

\begin{proof}[Proof of Lemma~$\ref{keyFail}$]
For ease of notation, we define
\begin{align*}
S_1 &:= p \sup_{\norm{y}_2 = 1} \left|\sum_{i=1}^n \sum_{j \le n, j \neq i} Q_{i,j}x_ix_jy_iy_j \right|\\
S_2 &:= \sqrt{p} \sqrt{\sum_{i=1}^{n} \sum_{j=1}^{n} Q^2_{i,j}x_i^2 x_j^2}.
\end{align*}

Our goal is to calculate $U_p = \norm{S_1 + S_2}_{p}$. We make use of the following upper and lower bounds on $\norm{S_1 + S_2}_{p}$:
\begin{equation}
\label{tighttriangleineq}
\left|\norm{S_1}_{p} - \norm{S_2}_{p}\right| \leq \norm{S_1 - S_2}_{p} \leq \norm{S_1 + S_2}_{p} \leq \norm{S_1}_{p} + \norm{S_2}_{p}.
\end{equation}
\noindent In order to compute $\left|\norm{S_1}_{p} - \norm{S_2}_{p}\right|$ and $\norm{S_1}_{p} + \norm{S_2}_{p}$, we first compute $\norm{S_1}_{p}$ and $\norm{S_2}_{p}$. For our choice of $x$, notice
\begin{align*}
\norm{S_1}_{p} &\simeq p \norm{\sup_{\norm{y}_2 = 1} \left| Q_{1,2}y_1y_2 \right|}_{p} \simeq  p \norm{ Q_{1,2}}_{p} \\
\norm{S_2}_{p} &\simeq \sqrt{p}  \norm{\sqrt{Q^2_{1,2}}}_{p} 
= \sqrt{p} \norm{Q_{1,2}}_{p}.
\end{align*}
From these bounds, coupled with $\eqref{tighttriangleineq}$, it follows that $\norm{U}_{p} \simeq p\norm{Q_{1,2}}_{p}$ as desired.
\end{proof}
We now show Lemma~$\ref{HWFailbound}$ follows from Lemma~$\ref{keyFail}$ and Proposition~$\ref{eqasym}$. 
\begin{proof}[Proof of Lemma~$\ref{HWFailbound}$]
After applying Lemma~$\ref{keyFail}$, it suffices to calculate $\norm{Q_{1,2}}_p$. It follows from Proposition~$\ref{eqasym}$ that $\norm{Q_{1,2}}_{p} \simeq \norm{X}_{p}$ where $X$ is distributed as $\Bin(s, s/m)$. Now, the following calculation $\norm{X}_p$ for $p < s$ and $B = m/s^2 \le \frac{e^p}{p}$ follows from the lower and upper bounds of Lemma~$\ref{estimate}$ (Lata{\l}a's bound on moments of sums of i.i.d nonnegative random variables):
\[\norm{X}_{p} \simeq 
\begin{cases}
\frac{p}{\log Bp} & \text{ if } B \ge \frac{e}{p} \\
\frac{1}{B} & \text{ if } B < \frac{e}{p}
\end{cases}.\]
From this, Lemma~$\ref{HWFailbound}$ follows. 
\end{proof}
\end{document}